\documentclass[aps,prl,twocolumn,amssymb,superscriptaddress,showpacs]{revtex4-1}
\usepackage{float}
\usepackage{xcolor}
\usepackage{graphicx}
\usepackage{dcolumn}
\usepackage{bm}
\usepackage{hyperref}
\usepackage{color}
\usepackage{amsmath}
\usepackage{amsfonts}

\begin{document}

\title{Multiple types of topological fermions in transition metal silicides}

\author{Peizhe Tang}
\affiliation{Department of Physics, McCullough Building, Stanford University, Stanford, California 94305-4045, USA}

\author{Quan Zhou}
\affiliation{Department of Physics, McCullough Building, Stanford University, Stanford, California 94305-4045, USA}

\author{Shou-Cheng Zhang}
\affiliation{Department of Physics, McCullough Building, Stanford University, Stanford, California 94305-4045, USA}
\affiliation{Stanford Institute for Materials and Energy Sciences, SLAC National Accelerator Laboratory, Menlo Park, California 94025, USA.}


\begin{abstract}
Exotic massless fermionic excitations with non-zero Berry flux, other than Dirac and Weyl fermions, could exist in condensed matter systems under the protection of crystalline symmetries, such as spin-1 excitations with 3-fold degeneracy and spin-3/2 Rarita-Schwinger-Weyl fermions. Herein, by using \emph{ab initio} density functional theory, we show that these unconventional quasiparticles coexist with type-I and type-II Weyl fermions in a family of transition metal silicides, including CoSi, RhSi, RhGe and CoGe, when the spin-orbit coupling (SOC) is considered. Their non-trivial topology results in a series of extensive Fermi arcs connecting projections of these bulk excitations on side surface, which is confirmed by (010) surface electronic spectra of CoSi. In addition, these stable arc states exist within a wide energy window around the Fermi level, which makes them readily accessible in angle-resolved photoemission spectroscopy measurements.
\end{abstract}


\maketitle

\emph{Introduction}.-- Three types of fermions play fundamental roles in our understanding of nature: Majorana, Dirac and Weyl \cite{pal2011dirac}. Much attention has been paid to looking for these fundamental particles in high energy physics during past few decades, whereas only signature of Dirac fermions is captured. Interestingly, the same movement comes up in the field of condensed matter physics \cite{armitage2017weyl}, and great achievements have been made in last few years. For example, the Majorana-like excitations are detected in superconducting heterostructures \cite{wang2012coexistence,nadj2014observation,WangJ2015,he2016chiral}; the Dirac \cite{WangZJ2012,WangZJ2013,tang2016dirac,liu2014stable,liu2014Na3Bi,xu2015observation} and Weyl \cite{WengHM2015,LvBQ2015,huang2015weyl,liu2016evolution,xu2015discovery,yang2015weyl,lv2015observation,SunY2015,soluyanov2015type,SunY2015MoTe2,WangZJ2016MoTe2,AutesG2016,deng2016experimental,huang2016spectroscopic,chang2016prediction,Koepernik2016,chang2016strongly} fermions are observed in some compounds. These quasiparticles in solid states are not only important for basic science, but also show great potential for practical applications on new devices \cite{SunY2016SpinHall,wang2016gate}.

\begin{figure}
\centerline{\includegraphics[clip,width=0.8\linewidth]{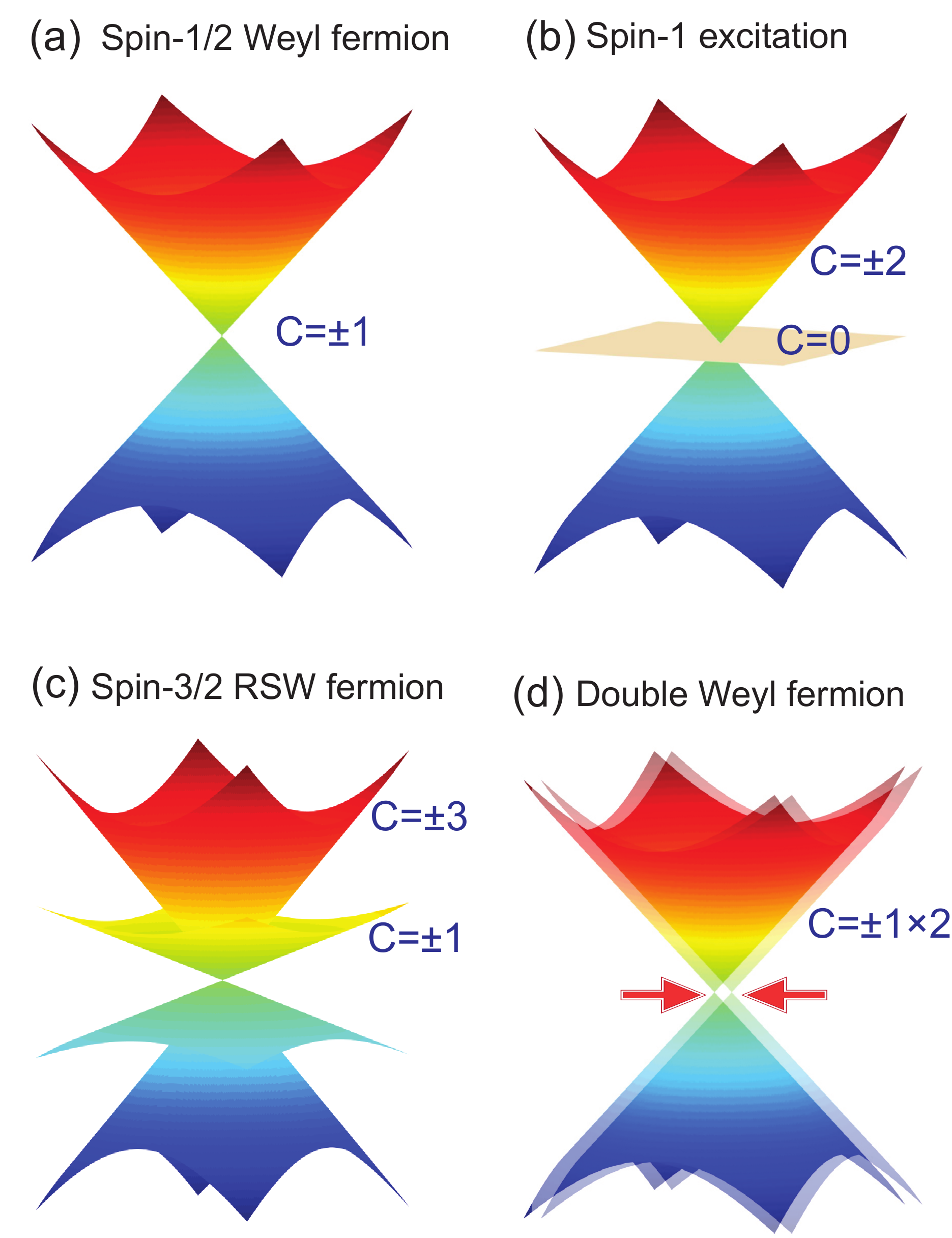}}
\caption{(Color online) Energy dispersions for multiple types of topological fermions. (a) The Weyl fermion with \textbf{S}=1/2. (b) The excitation with \textbf{S}=1. (c) The Raita-Schwinger-Weyl fermion with \textbf{S}=3/2. (d) The double Weyl fermion. The red arrows indicate that two energy crossings should be at the same point for the double Weyl fermion. Chern numbers for upper and lower bands are marked in blue for topological fermions.}
\label{fig:1}
\end{figure}

\begin{figure*}
\centerline{\includegraphics[clip,width=0.85\linewidth]{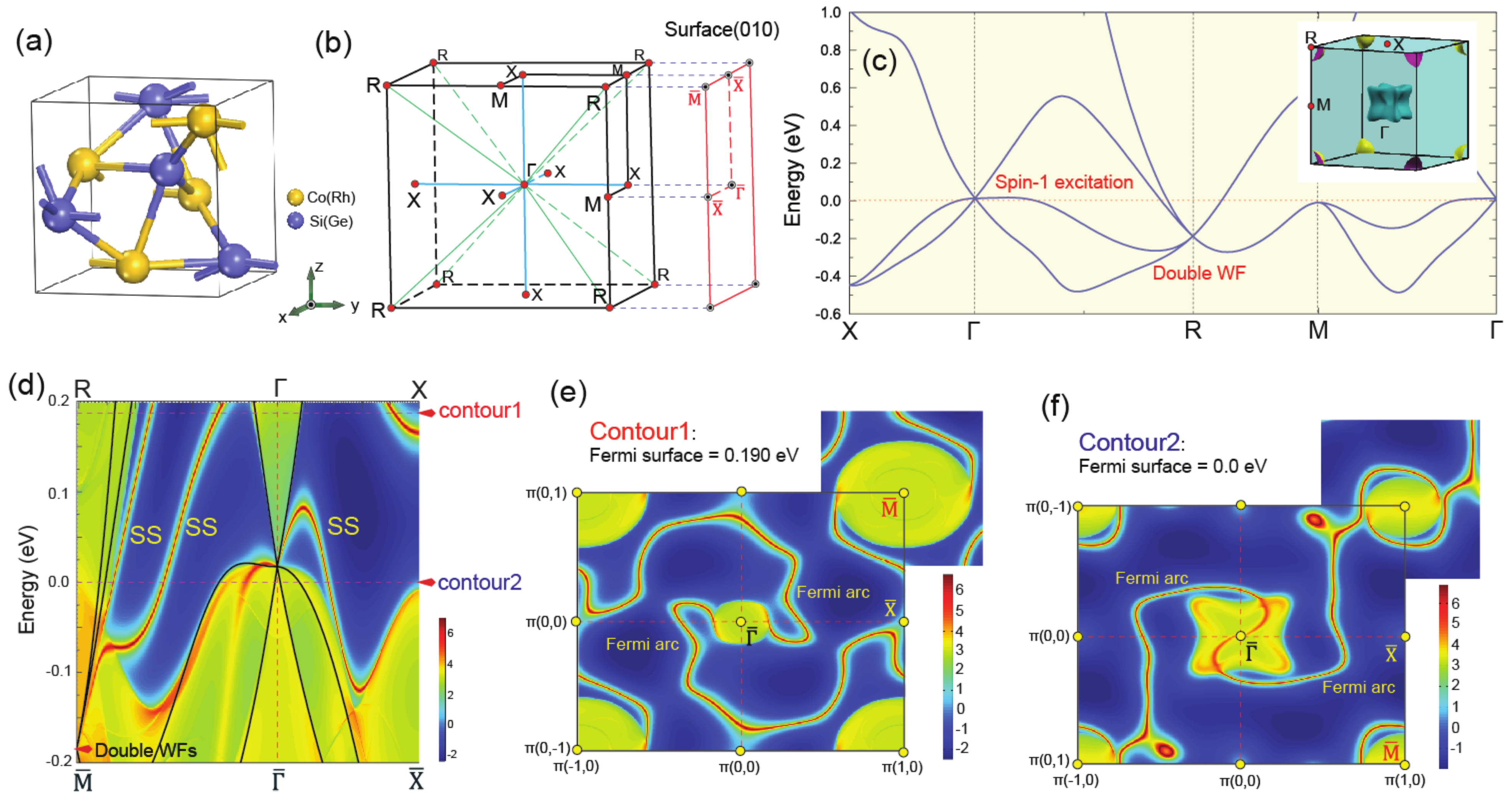}}
\caption{(Color online) The crystalline lattice structure, Brillouin zone (BZ) and electronic properties for CoSi without SOC. (a) The lattice structure and (b) BZ for CoSi-family. The projected BZ of (010) surface is marked in red lines. (c) The bulk band structure for CoSi along high symmetry lines. Inset is 3D Fermi surface at the calculated Fermi level. (d) The corresponding electronic spectra for (010) surface. Bulk bands along lines of R-$\Gamma$ and $\Gamma$-X are plotted in black, in which the line along R-$\Gamma$ are re-scaled to fit the distance between $\bar{\rm M}$ and $\bar{\Gamma}$ points. (e, f) The Fermi surface contours on (010) surface at different energies. The calculated Fermi level is set to be zero.}
\label{fig:2}
\end{figure*}

Because symmetries in condensed matter physics are usually much lower than the Poincar$\rm \acute{e}$ symmetry in high energy physics, quasiparticles in solid states are less constrained such that various new types of fermionic excitations are predicted to exist in 3D lattices \cite{bradlyn2016beyond}. Among these allowed by space group (SG) symmetries are spin-1 and spin-3/2 massless fermionic excitations, besides the well-known spin-1/2 case, namely the Weyl fermion. All of these massless quasiparticles can be described by the low energy Hamiltonian in a unified manner to the linear order of momentum
\begin{gather}
H = \delta\bm{k}\cdot \bm{S}
\end{gather}
where $\delta{\bm k} = \bm{k} - \bm{k}_0$ is the momentum deviation from the crossing point $\bm{k}_0$, and $\bm{S}$ stands for the matrices for pseudo-spin degree of freedom that satisfy $[\bm{S}_i, \bm{S}_j] = i\epsilon_{ijk} \bm{S}_k$. The definite helicity can be assigned to each energy band of $H$, and it is related to the non-vanishing Chern number for one surface enclosing the crossing point. These band crossings behave as monopoles of the Berry flux. For example, the Weyl fermion takes $2\times 2$ Pauli matrices and holds a two-fold degeneracy, its crossing point carries a topological charge $\pm 1$ (see Fig. \ref{fig:1}(a)). As a generalization, the spin-1 excitation takes $3 \times 3$ spin matrices and holds a three-fold degeneracy (see Fig. \ref{fig:1}(b)), its crossing point carries a topological charge $\pm 2$ because of no contribution from the middle band with the helicity 0 \cite{bradlyn2016beyond,Manes2012}. Spin-3/2 excitations are named as Rarita-Schwinger-Weyl (RSW) fermions \cite{Rarita1941,LiangL2016,bradlyn2016beyond,Ezawa2016}, and $\bm{S}$ takes $4\times 4$ spin matrices. Its Fermi surface has to cross two bands near the crossing point, for example, the helicity 3/2 band with the Berry phase $\pm 3$ and the helicity 1/2 band with the Berry phase $\pm 1$ (see Fig. \ref{fig:1}(c)). So a spin-3/2 RSW fermion acts as a monopole with topological charge $\pm 4$. Additionally, the time-reversal (TR) symmetry could lead to a non-trivial doubling of these excitations in some situations, which gives birth to double Weyl fermions possessing topological charge $\pm 2$ (see Fig. \ref{fig:1}(d)) \cite{sm1,geilhufe2016data}, and double spin-1 excitations with 6-fold degeneracy \cite{bradlyn2016beyond}.

In this work, we explore electronic properties for a family of transition metal silicides (CoSi-family) based on \emph{ab initio} calculations. Its bulk states host all of exotic fermionic quasiparticles mentioned above under different conditions. To our best knowledge, it is the first attempt to incorporate multiple types of topological fermions in a solid compound that has been synthesized experimentally \cite{shinoda1972magnetic,geller1954crystal,takizawa1988high,larchev1982polymorphism}. Around the Fermi level, the conductance in CoSi-family compounds are mainly contributed by these unconventional excitations. Therefore the signature related to non-trivial topology is possible to be observed in transport measurements clearly. On the other hand, because these fermionic excitations with different topological charges reside in (or near) either the center or the corner of the Brillouin zone (BZ), the surface states that connect their projections, emerge extensively on the side surface, in sharp contrast to most known topological semimetals whose Fermi arcs only live within small regions \cite{LvBQ2015,liu2016evolution,SunY2015,yang2015weyl,lv2015observation,SunY2015MoTe2,WangZJ2016MoTe2,deng2016experimental,huang2016spectroscopic,chang2016prediction,soluyanov2015type}. The large extent of arc states offers a great opportunity to fully investigate the non-trivial surface states in experiments.

\emph{Methods}.-- The \emph{ab initio} calculations were carried out in the framework of density functional theory with the projector augmented wave method \cite{PhysRevB.50.17953,PhysRevB.59.1758}, as implemented in the Vienna \textit{ab initio} simulation package \cite{PhysRevB.54.11169}. Plane wave basis set was used with a kinetic energy cutoff of $\mathrm{300~eV}$. The lattice constant was from previous experimental observations \cite{shinoda1972magnetic}. Atomic positions in one cubic lattice were allowed to fully relax until residual forces were less than $1\times 10^{-3}~\mathrm{eV/\AA}$. The Monkhorst-Pack $k$ points were $9\times 9\times 9$, and SOC was included in self-consistent electronic structure calculations. The maximally localized Wannier functions \cite{mostofi2008wannier90} were constructed to obtain the tight-binding Hamiltonian for the Green's function method, which was used to calculate the surface electronic spectra \cite{Wu2017}.

\begin{figure}[ht]
\includegraphics[clip,width=\linewidth]{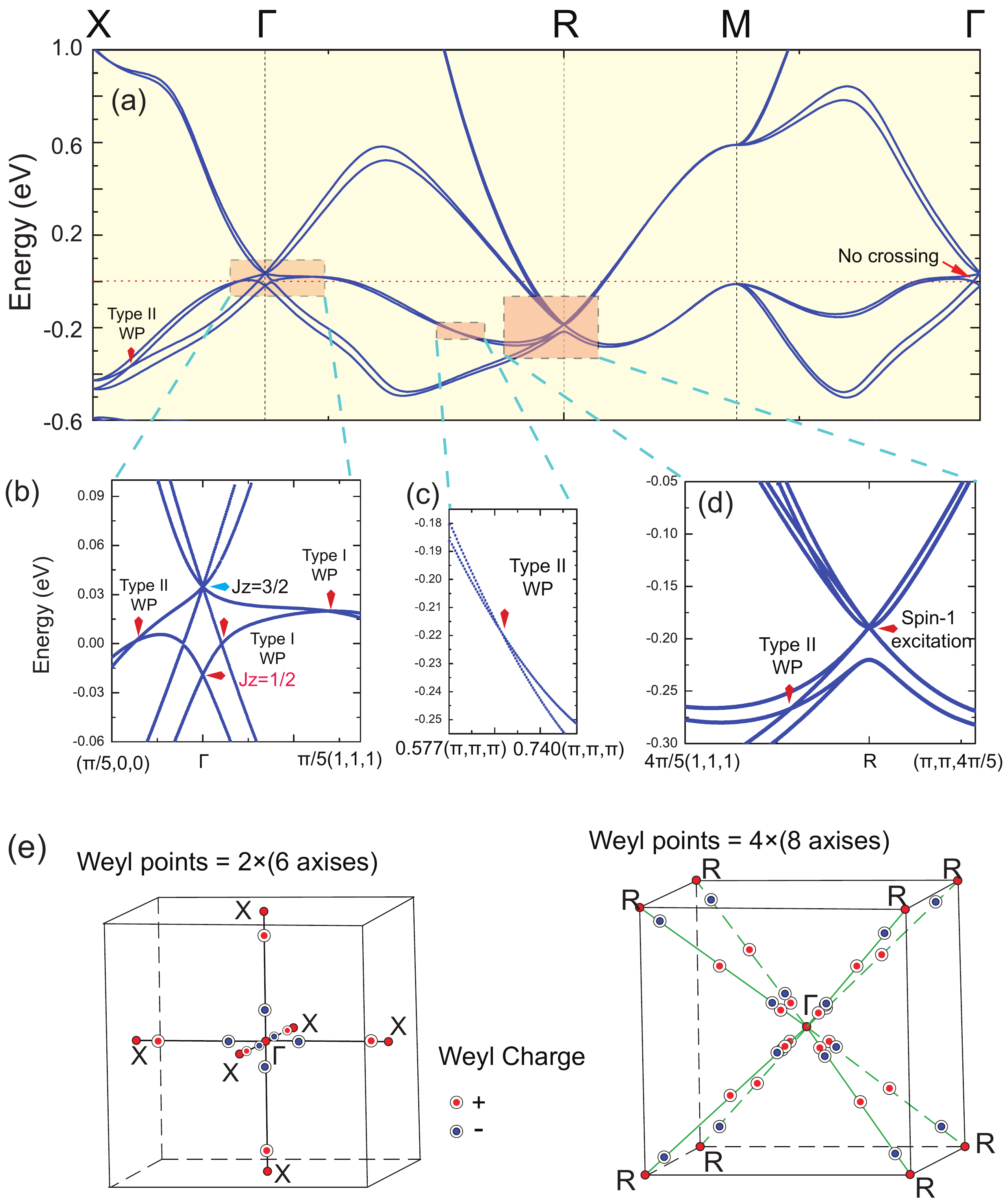}
\caption{(Color online) Electronic structures for CoSi with SOC. (a) The bulk band structure for CoSi along high symmetry lines. (b-d) Zooming-in of bands in regions marked by red boxes. Weyl fermions are marked by the red arrows (including type-I and type II). The Raita-Schwinger-Weyl fermion is marked by blue arrow. The calculated Fermi level is set to be zero.  (e) Schematics of the distribution of Weyl fermions along lines of $\Gamma$-X and $\Gamma$-R around the Fermi level. The red (blue) dots stand for the Weyl points with positive (negative) topological charge.}
\label{fig:3}
\end{figure}

\emph{Lattice structure of CoSi-family}.-- The family of transition metal silicides includes CoSi \cite{shinoda1972magnetic}, RhSi \cite{geller1954crystal}, CoGe \cite{takizawa1988high} and RhGe \cite{larchev1982polymorphism}, which have been synthesized experimentally. All of them have the same cubic lattice with space group P$2_{1}$3 (No.198) and similar electronic properties. For convenience, we focus on CoSi, a prototype in this family, in the following discussion. Its lattice structure is shown in Fig. \ref{fig:2}(a), each Si site is bonded by four nearest neighboring Co atoms in one unitcell. The corresponding BZ is shown in Fig. \ref{fig:2}(b), TR invariant points are plotted by red dots. The SG 198 has 12 symmetry operations, which can be generated by three of them: one 3-fold rotation symmetry along (111) axis, two 2-fold screw symmetries along $z$ and $x$ axis. Therefore, the lattice of CoSi has three 2-fold and four 3-fold rotation or screw axises totally.

\emph{Electronic structures without SOC}.-- Figure \ref{fig:2}(c) demonstrates the calculated electronic structure and the Fermi surface for CoSi bulk without SOC. Although this compound contains transition metal, no magnetization is observed in our calculations, the whole system guarantees TR symmetry. This result is similar to the case in FeSi with SG 198 \cite{JKuler2013}. At the Fermi level, electronic states are only contributed by hole pockets at the $\Gamma$ point and electron pockets at the R point. For each physical spin, a gapless point with 3-fold degeneracy is observed above the Fermi level at the center of BZ, which is stabilized by crystal symmetries \cite{SuppMater}. Its low energy physics can be described by spin-1 excitations shown in Fig. \ref{fig:1}(b), whose crossing point is a monopole possessing topological charge +2. At the R point, we found a band crossing with 4-fold degeneracy below the Fermi level, which is a double Weyl fermion with the Chern number -2 \cite{sm1}. So the total Chern number is zero for whole Fermi surface in CoSi bulk, which is consistent with no-go theorem \cite{nielsen1981no}.

Due to non-trivial topology possessed by hole and electron pockets in the bulk, the Fermi arc surface states can be observed on the side surface. The electronic spectra for (010) surface is shown in Fig. \ref{fig:2}(d). We can see that topological surface states (marked by SS) emerge from projections of spin-1 excitation and double Weyl fermion at $\Gamma$ and R points, which are stable in a large energy window. Figs. \ref{fig:2}(e) and (f) demonstrate the Fermi surface contours on (010) surface at different energies, in which two Fermi arcs connect states at $\bar{\rm \Gamma}$ and $\bar{\rm M}$ points. Especially for the contour at the Fermi level (see Fig. \ref{fig:2}(e)), Fermi arcs emerge from the $\bar{\rm \Gamma}$ point directly, which indicates that the middle flat band in spin-1 excitations does not carry topological charge (see Fig. \ref{fig:1}(b)).

\begin{figure*}[ht]
\centerline{ \includegraphics[clip,width=0.95\linewidth]{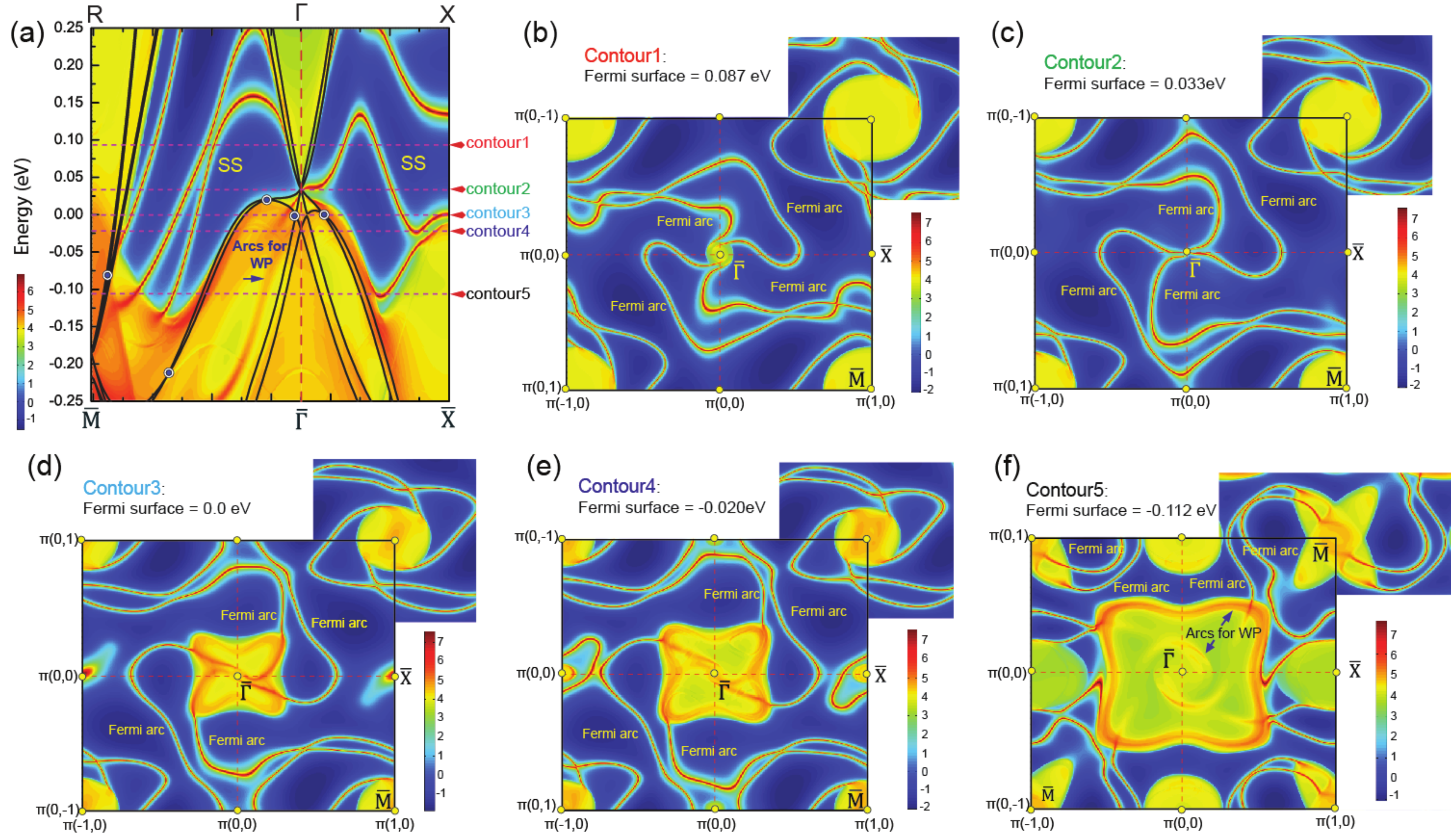}}
   \caption{(Color online) The projected electronic spectra for CoSi with SOC. (a) The electronic spectra along high symmetry lines projected to (010) surface. Bulk bands along lines of R-$\Gamma$ and $\Gamma$-X are plotted as black lines, in which the line along R-$\Gamma$ are re-scaled to fit the distance between $\bar{\rm M}$ and $\bar{\Gamma}$ points. Type-I and type-II Weyl fermions are marked by blue dots. (b-f) The corresponding Fermi surface contours on (010) surface at different energies. The calculated Fermi level is set to be zero.}
\label{fig:4}
\end{figure*}

\emph{Electronic structures with SOC}.-- Figure \ref{fig:3} shows the bulk band structures for CoSi with SOC. Due to the absence of inversion symmetry in SG 198, the SOC term lifts the degeneracy at arbitrary non-TR invariant point, except for states on boundaries of 3D BZ whose double degeneracy is protected by TR and non-symmorphic screw symmetries \cite{bradlyn2016beyond}. At the center of BZ, the 6-fold degeneracy point is splited by SOC into a 2-fold degeneracy point and a 4-fold degeneracy point, which correspond to a Weyl fermion and a spin-3/2 RSW fermion with topological charge +4 \cite{SuppMater} respectively. The 4-fold degeneracy originates from the TR forced doubling of the underlying two-dimensional irreducible representation of the symmetry group \cite{bradley2010mathematical}. And the existence of the spin-3/2 RSW fermion in a compound with SG 198 is beyond previous studies on unconventional quasiparticles \cite{bradlyn2016beyond}. Meanwhile, a crossing point with 6-fold degeneracy is found at R point. It is realized as a TR doubling of spin-1 excitations protected by non-symmorphic symmetries \cite{bradlyn2016beyond}, and its total topological charge is -4.

At the same time, we found CoSi can host type-I and type-II Weyl fermions along symmetry invariant axises. For bands below the gapless RSW point (see Fig. \ref{fig:3}), six pairs of type-II Weyl fermions exist along $\Gamma$-X lines, each pair has opposite chiral charges. And four Weyl fermions are observed on each line of $\Gamma$-R that is invariant under 3-fold rotation or screw symmetries. Two of them are type-I, the others are type-II. In total, 32 Weyl points exist along these 3-fold rotation or screw axises (see Fig. \ref{fig:3} (e)), and the sum of their topological charges is zero. Similar to previous discussions \cite{FangC2012,tsirkin2017composite}, these Weyl fermions in CoSi are generated by the crossing of states with different eigenvalues of rotation or screw symmetries.

In order to demonstrate exotic physics of topological fermions in CoSi with SOC, we explore its electronic spectra on (010) surface. The calculated results are shown in Fig. \ref{fig:4}. Non-trivial surface states that related to excitations beyond Dirac and Weyl models can be observed clearly on the side surface, which emerge from projections of bulk states at the $\bar{\Gamma}$ point and end at those around the $\bar{\rm M}$ point. In Figs. \ref{fig:4}(b)-(f), we show Fermi surface contours at different energies, all of them are above the gapless point with 6-fold degeneracy in energy scale. In contour 1 (see Fig. \ref{fig:4}(b)), states around the $\bar{\Gamma}$ point are projected from bulk conduction bands with helicities of 3/2 and 1/2 in the RSW fermion, their total topological charge is +4. So four Fermi arcs are observed around the $\bar{\Gamma}$ point on (010) surface. When the energy cuts the crossing point in the RSW fermion (see Fig. \ref{fig:4}(c)), Fermi arcs emerge from the $\bar{\Gamma}$ point directly, indicating the non-trivial topology carried by the RSW fermion. Furthermore, with lower energy, projections at the $\bar{\Gamma}$ point are from states with 3/2 helicity in RSW fermion below its 4-fold gapless point and states in Weyl fermion at the $\Gamma$ point. Their total Chern number still is 4, four Fermi arcs are observed in Figs. \ref{fig:4}(d)-(f). Similar to cases in TaAs \cite{LvBQ2015,liu2016evolution,SunY2015,yang2015weyl,lv2015observation} and Mo$\rm Te_2$ \cite{SunY2015MoTe2,WangZJ2016MoTe2,deng2016experimental,huang2016spectroscopic,chang2016prediction,soluyanov2015type}, Fermi arcs contributed by type-I and type-II Weyl fermions along rotation or screw axises are coupled with bulk states strongly, which makes their signatures hard to be distinguished. Herein, we found the possible signature of Fermi arcs contributed by type-II Weyl fermions in Fig. \ref{fig:4}(f) and the Lifitiz transition for Fermi arc surface states.

\emph{Conclusion}.-- By using first principles calculations, we predict that bulk states of CoSi-family compounds host spin-1 excitations, double Weyl fermions, spin-3/2 RSW fermions, type-I and type-II Weyl fermions in cases without and with SOC. The corresponding extensive Fermi arcs are observed on (010) surface clearly. Different from previous found topological semimetals, our systems support topological features in a large energy window around the Fermi level. We expect that these electronic signatures can be observed via angle-resolved photoemission spectroscopy directly. When magnetic field is applied to this system to break TR symmetry, these quasiparticles will split to multiple Weyl fermions \cite{bradlyn2016beyond}, and the anomalous magnetoresistance and chiral anomaly can be detected by transport measurements. Therefore, this work not only identifies a series of desired robust topological semimetal candidates, but also provides an ideal platform to explore exotic physical phenomena and future device applications.

We acknowledge the Department of Energy, Office of Basic Energy Sciences, Division of Materials Sciences and Engineering, under contract DE-AC02-76SF00515, and FAME, one of six centers of STARnet, a Semiconductor Research Corporation program sponsored by MARCO and DARPA.

\emph{Note added}.-- After completing our manuscript, we become aware of a preprint aiming at double Weyl points in phonon dispersion for transition-metal silicides MSi (M=Fe, Co, Mn, Re, Ru) with SG 198. \cite{ZhangTT2017}


%

\end{document}